\documentclass[aps,prd,preprint,12pt,secnumarabic,tightenlines,amsmath,amssymb]{revtex4}
\usepackage{epsfig}
\usepackage{graphicx}
\usepackage{bm}

\begin{document}
\begin{center}
\Large \bf Cosmology calculations almost without general relativity \\

\end{center}

\begin{center}
Thomas F. Jordan\\
Physics Department \\
University of Minnesota\\
Duluth, Minnesota 55812\\
tjordan@d.umn.edu\\
\end{center}

$\newline$

\begin{abstract}
The Friedmann equation is derived for a Newtonian universe. Changing
mass density to energy density gives exactly the Friedmann equation
of general relativity. Accounting for work done by pressure then
yields the two Einstein equations that govern the expansion of the
universe. Descriptions and explanations of radiation pressure and
vacuum pressure are added to complete a basic kit of cosmology
tools. It provides a basis for teaching cosmology to undergraduates
in a way that quickly equips them to do basic calculations. This is
demonstrated with calculations involving: characteristics of the
expansion for densities dominated by radiation, matter, or vacuum;
the closeness of the density to the critical density; how much
vacuum energy compared to matter energy is needed to make the
expansion accelerate; and how little is needed to make it stop.
Travel time and luninosity distance are calculated in terms of the
redshift and the densities of matter and vacuum energy, using a
scaled Friedmann equation with the constant in the curvature term
determined by matching with the present values of the Hubble
parameter and energy density. General relativity is needed only for
the luminosity distance, to describe how the curvature of space,
determined by the energy density, can change the intensity of light
by changing the area of the sphere to which the light has spread.
Thirty-one problems are included.

\end{abstract}
\maketitle
\section{Introduction}
 To understand cosmology, it is necessary to work with equations
that come from general relativity.  That does not mean a student
needs a course in general relativity to get started.  Substantial
working knowledge of cosmology can be built on simple undergraduate
physics. The Friedmann equation is the key.  It is one of the two
Einstein equations of general relativity that govern the expansion
of the universe.  The other can be easily understood; it accounts
for work done by pressure as the universe expands.

 The  Friedmann equation can  be  derived for a Newtonian universe
where motion is nonrelativistic  and  gravity  is attraction between
$\text{masses.} ^{\mathbf{1}}$ In this case, the Friedmann equation
describes conservation of nonrelativistic kinetic energy plus
gravitational potential energy for the motion of a galaxy in the
expansion.  To move to relativity we use the equivalence of mass and
energy and change mass density to energy density.  What we need to
know about general relativity at this step is that it gives exactly
the same Friedmann equation in all cases.  The foundation is
curvature of space rather than gravitational energy, but the result
is the same. There are no correction terms.  The only change we have
to make is to interpret the density in the Friedmann equation as
that of energy rather than mass and include all forms of energy in
it.

 That is reviewed and developed here.  Descriptions and
explanations of radiation pressure and vacuum pressure are added to
complete a basic kit of cosmology tools.  It fills in the
mathematics left out of descriptive reviews $^{\mathbf{2}}$ and
provides a basis for teaching cosmology to undergraduates in a way
that quickly equips them to do basic calculations.  Application of
the equations is demonstrated here with calculations that: determine
characteristics of the expansion for cases where the energy density
is dominated by radiation, matter, or vacuum; present and answer the
question how the energy density got so close to the critical
density; find how much vacuum energy compared to matter energy is
needed to make the expansion accelerate; and find how little is
needed to make it stop.

 The travel time and luminosity distance for light from a
distant source are calculated in terms of the redshift and the
densities of matter and vacuum energy.  In the scaled Friedmann
equation that is used for this, the constant in the curvature term
is determined by matching with the present values of the Hubble
parameter and energy density and is found to be a function of them.
General relativity is not needed for that, but it is needed to
describe how the curvature of space can change the area of the
sphere of light and, thereby, the intensity of the light spread over
that sphere, and hence the luminosity distance used in Hubble plots.
In particular, general relativity gives the curvature in terms of
the energy density.  Thirty-one problems, most of them easy, explore
some of these topics in more detail and extend the range of
application. 

\section{Friedmann Equation}
 Two basic equations of cosmology govern the expansion of the
universe.  They can be understood as statements about energy.  One
says there is a constant like the sum of kinetic and gravitational
energy for the motion of a galaxy in the expansion.  The other
equation accounts for the work done by pressure as the universe
expands.

 The expansion of the universe is observed as motion of galaxies away
from each other.  The universe is like an expanding gas, but the
units are galaxies; an individual galaxy does not expand.  The
Hubble law describes what is $\text{observed.}^{\mathbf{3}}$ The
speeds of galaxies moving away from us are proportional to their
distances from us.  Galaxies at distances $R$
 are moving away from us with average speed
\begin{equation}
\frac{dR}{dt} = HR \tag{2.1}
\end{equation}
where $H$
 is the Hubble parameter.
 The Hubble law is what we expect to observe in a spatially isotropic
uniformly expanding $\text{universe.}^{\mathbf{3}}$ We assume it
would be the same for observers at any other location.  We assume
the universe is in fact homogeneous, the same everywhere, at any
given time.  Right now, that can be true only at scales larger than
clusters of galaxies, but the isotropy of the cosmic radiation is
evidence that at earlier times the universe was in fact very
$\text{homogeneous.}^{\mathbf{4}}$ Gravity pulls the galaxies
together and slows the expansion of the universe.  If distances are
measured
 from a typical galaxy, which could be at any location, the force of gravity on a galaxy at distance $R$
 coming from the mass of homogeneous universe inside the sphere of
radius $R$ is the same as if all that mass were at the center of the
sphere. There is no force of gravity from the spherical shells of
homogeneous universe outside that sphere.  The kinetic plus
gravitational potential energy for the motion of a galaxy of mass
$m$
 in the expansion is therefore
\begin{equation}
E = \frac{1}{2} m (\frac{dR}{dt})^{2} - \frac{GMm}{R} \tag{2.2}
\end{equation}
where $G$
 is Newton's gravitational constant,
\begin{equation}
M = \frac{4}{3} \pi R^{3} \rho \tag{2.3}
\end{equation}
is the mass inside the sphere of radius $R,$ and $\rho$ is the
density of mass in the universe.  Using the Hubble law, Eq. (2.1),
we get
\begin{equation}
\frac{2E}{mR^{2}} = H^{2} - \frac{8}{3} \pi G \rho. \tag{2.4}
\end{equation}
At a given time in our homogeneous universe, $H$ and $\rho$ are
constant throughout space.  That means $2E / mR^{2}$ is the same for
all galaxies.  In particular, the negative, zero, or positive
character of $E$
 is the same for all galaxies.

  If $E$
 is not zero, we can choose units and find a time
$t_1$ so that $| 2E / mR(t_1)^{2}|$ is 1. Then
\begin{equation}
\frac{( dR/dt)^{2}}{R^{2}} - \frac{8}{3} \pi G \rho = -
\frac{kR(t_1)^{2}}{R^{2}} \tag{2.5}
\end{equation}
where $k = -2E / mR(t_1)^{2}$ is 1, 0 or -1 depending on whether $E$
 is negative, zero, or positive.  The constant
$R(t_1)^{2}$ is the magnitude of $2E/m$ . It stays constant while $R
, dR/dt , H \text{ and } \rho$ change in time.  (This constant will
be handled in a completely different way in Sections IX and X where
applications require its use. Meanwhile we will see how much can be
done without knowing anything more about it.)

  Now we invoke $E = m c^{2}$ and let $\rho$
 be the energy density of the universe.  We use units for which
$c$ is 1.  Mass and energy are equivalent.  We assume, as in general
relativity, that all energy creates gravity. With $\rho$ the energy
density, our Eq. (2.5) is the Friedmann equation obtained from
general relativity, where $k$ is the constant 1, 0, or -1 that
indicates whether the curvature of space is positive, zero, or
$\text{negative.}^{\mathbf{5}}$

  In our Newtonian equations, $E$
 is constant, so
$k$ is constant.  We have derived the Friedmann equation for the
case of a Newtonian universe.  General relativity says the Friedmann
equation holds with constant $k R(t_1)^{2}$ in all cases.  Of course
general relativity provides a derivation of the Friedmann equation
that is altogether more satisfactory.  But Newtonian physics comes
remarkably close.  It gives us an equation that is exactly the
Friedmann equation where Newtonian physics applies and with
reinterpretation becomes exactly the Friedmann equation in all
cases.  It does not require correction terms.

\section{Einstein Equations}

  The Friedmann equation is one of the two Einstein equations
of general relativity for the homogeneous universe.  The other just
adds that as a volume $V$ of the universe expands by an amount $dV$,
 the pressure $p$ in the volume does work $p dV,$
 which decreases the energy in the volume by that amount. So for a
sphere of radius $R$,
\begin{equation}
d(\rho \frac{4}{3} \pi R^{3}) = - p d(\frac{4}{3} \pi R^{3} )
\tag{3.1}
\end{equation}
which implies that
\begin{equation}
R \frac{d \rho}{dt} + 3( \rho +p ) \frac{dR}{dt} = 0.  \tag{3.2}
\end{equation}
We can write the Friedmann Eq. (2.5) as
\begin{equation}
(\frac{d R}{dt})^{2} = \frac{8}{3} \pi G \rho R^{2} - k R(t_1)^{2}.
\tag{3.3}
\end{equation}
Taking the time derivative of this, using Eq. (3.2) for $R d \rho /
dt $ , yields
\begin{equation}
\frac{d^{2} R}{dt^{2}} = -\frac{4}{3} \pi G ( \rho + 3p ) R.
\tag{3.4}
\end{equation}
These last two Eqs. (3.3) and (3.4) are the two basic equations of
cosmology obtained from general $\text{relativity.}^{\mathbf{5}}$

\section{Density and Pressure}
  The pressure depends on the nature of energy.
The energy density can come from matter, radiation, vacuum, or
combinations of these.  Each gives a different pressure.

  The energy density of matter does not produce pressure that
affects the expansion of the universe.  Galaxies are not colliding
and creating pressure the way molecules in a gas do. In the early
universe, the energy density was mostly from radiation. The density
and pressure of radiation are related by
\begin{equation}
p_\gamma = \frac{1}{3} \rho_\gamma. \tag{4.1}
\end{equation}
That is shown in Appendix A.

  The work done by each constituent decreases the energy of
that constituent as the universe expands,  so Eq. (3.1) holds
separately for each constituent:
\begin{equation}
\frac{d}{dt} \rho_i R^3 = - p_i \frac{d}{dt} R^3 \tag{4.2}
\end{equation}
where $i \text{ can be } m, \gamma \text{ or } v$ for matter,
radiation, or vacuum.  For matter, $p_m$ is zero, so
\begin{equation}
\frac{d}{dt} \rho_m R^3 = 0. \tag{4.3}
\end{equation}
The energy in a sphere of matter does not change as the sphere
expands; it just spreads out.  The time dependence of $\rho_m$ is
tied to that of $R$ by the relation
\begin{equation}
\rho_m \propto \frac{1}{R^{3}}. \tag{4.4}
\end{equation}

  For radiation, Eqs. (4.1) and (4.2) give
\begin{equation}
\frac{d}{dt} \rho_\gamma R^4 = \rho_\gamma R^3 \frac{dR}{dt} + R
\frac{d}{dt} \rho_\gamma R^3 \notag
\end{equation}
\begin{equation}
= \rho_\gamma R^3 \frac{dR}{dt} - R \frac{1}{3} \rho_\gamma
\frac{d}{dt} R^3 = 0 \tag{4.5}
\end{equation}
so the time dependence of $\rho_\gamma$ is determined by
\begin{equation}
\rho_\gamma \propto \frac{1}{R^4}. \tag{4.6}
\end{equation}

 As the universe expands, the wavelengths $\lambda$ of
radiation expand in proportion to distances $R$ . Since the
wavelengths $\lambda$
 and frequencies
$\nu$ of radiation are related by
\begin{equation}
\nu \lambda = c \tag{4.7}
\end{equation}
and the energy of a photon is
\begin{equation}
E = h \nu, \tag{4.8}
\end{equation}
and since the average energy of a photon of radiation in equilibrium
at temperature $T$ $\text{is}^{\mathbf{6}}$
\begin{equation}
\bar{E} \approx 2.7 k T \tag{4.9}
\end{equation}
where $h \text{ and } k$ are the Planck and Boltzmann constants,
which we assume do not change, we see that the way wavelengths,
frequencies, photon energies and temperature change as the universe
expands is determined by
\begin{equation}
\lambda \propto R \notag
\end{equation}
\begin{equation}
E \propto \nu \propto \frac{1}{R} \notag
\end{equation}
\begin{equation}
T \propto \bar{E} \propto \frac{1}{R}. \tag{4.10}
\end{equation}
The number of photons does not change as the universe expands.  The
number of photons per unit volume $n_\gamma$ decreases as the volume
expands so
\begin{equation}
n_\gamma \propto \frac{1}{R^{3}}. \tag{4.11}
\end{equation}
The energy per unit volume of the radiation is
\begin{equation}
\rho_\gamma = n_\gamma \bar{E}. \tag{4.12}
\end{equation}
This explains Eq. (4.6). For radiation in equilibrium at temperature
$T,$
 the Stefan-Boltzmann-Planck formulas
$\text{give}^{\mathbf{6}}$
\begin{equation}
n_\gamma \propto T^{3} , \qquad  \qquad \rho_\gamma \propto T^{4}
\tag{4.13}
\end{equation}
which agree with Eqs. (4.6), (4.10) and (4.11).  It all fits
together.

  The energy density of vacuum does not change:
\begin{equation}
\frac{d}{dt} \rho_v = 0. \tag{4.14}
\end{equation}
That means
\begin{equation}
\frac{d}{dt} \rho_v R^{3} = \rho_v \frac{d}{dt} R^3 \tag{4.15}
\end{equation}
so Eq. (4.2) implies
\begin{equation}
p_v = - \rho_v . \tag{4.16}
\end{equation}

 There are layers of understanding this.  Superficially we can say
the energy density of vacuum does not change because there is
nothing there to change.  To probe more deeply, we remember that the
vacuum is supposed to be Lorentz invariant.  In a homogeneous
universe, the energy density is constant throughout space.  The
spatial gradient of the energy density is zero.  Lorentz
transformations mix space and time derivatives, so for Lorentz
invariance the time derivative of the energy density must be zero
too. We can explore this further by considering the energy-momentum
tensor.  It involves both energy density and pressure.  Its Lorentz
invariance requires that they are related by Eq. (4.16).  That is
shown in Appendix B.

 Vacuum energy and pressure have the same effect as a cosmological
constant in the equations of general relativity. Einstein used the
symbol $\Lambda$ for the cosmological constant and it is still often
used to label vacuum energy and pressure.
$\newline$

{\bf Problem 4.1.}  Suppose there is an energy density $\rho_w$
 and pressure
$p_w$
 related by
\begin{equation}
p_w = w \rho_w \notag
\end{equation}
with $w$ a constant.  Show that $\rho_w R^{3+3w}$ stays constant as
the universe expands so the time dependence of $\rho_w$ is
determined by
\begin{equation}
\rho_w \propto R^{-3-3w}. \notag
\end{equation}
Note that Eqs. (4.4), (4.6) and (4.14) are obtained as special cases
when $w$ is 0, 1/3 and -1.  A variety of hypotheses about contents
of the $ \text{universe}^{\mathbf{7}} $
 is covered by broad use of the name ``quintessence" and a
pressure/energy ratio $w$, particularly for $w$ less than $ -1/3$,
but also more generally for time-dependent as well as constant $w$.

\section{Time Dependence}

  Many features of cosmology can be explained with calculations
using the equations we have developed.  A few central examples are
worked out here. The equations yield solutions that describe both
expansion and contraction of the universe.  We will focus on
expansion.

  As the universe expands, the energy density of radiation
decreases faster, Eq. (4.6), than that of matter, Eq. (4.4), and the
energy density of vacuum does not decrease at all, Eq. (4.14).  At
early times, the energy density of radiation must have been
dominant.  If there is vacuum energy density, there will be a time
in the future when it is dominant.  In our universe, it appears that
for a time in between, the energy density of matter is important if
not dominant.

 The time dependence of the expansion of the universe is different
for the pressures of radiation, matter and vacuum, so it is
different in eras when different kinds of energy are dominant. To
see the simplest examples, suppose $k$ is zero or is negligible in
the Friedmann equation (as it is at early times when the term
involving $k$ is overwhelmed by that involving the energy density of
radiation). Then the Friedmann equation says
\begin{equation}
(\frac{dR}{dt})^{2} \propto \rho R^2. \tag{5.1}
\end{equation}
We consider simple cases where the universe contains only one
constituent, either radiation, matter, or vacuum.  For radiation,
from Eq. (4.6), we see that
\begin{equation}
\frac{dR}{dt} \propto \frac{1}{R} \tag{5.2}
\end{equation}
so
\begin{equation}
R \propto t^{\frac{1}{2}}. \tag{5.3}
\end{equation}
For matter, from Eq. (4.4), we get
\begin{equation}
\frac{dR}{dt} \propto \frac{1}{R^{\frac{1}{2}}} \tag{5.4}
\end{equation}
\begin{equation}
R \propto t^{\frac{2}{3}}. \tag{5.5}
\end{equation}
For vacuum, from Eq. (4.14), we get
\begin{equation}
\frac{dR}{dt} \propto R \tag{5.6}
\end{equation}
\begin{equation}
R \propto e^{H t} \tag{5.7}
\end{equation}
with constant $H = (dR/dt)/R.$ This exponential expansion produced
by vacuum energy is called inflation. If radiation, matter and
vacuum all contribute, the density and pressure are sums
\begin{equation}
\rho = \rho_\gamma + \rho_m + \rho_v \tag{5.8}
\end{equation}
\begin{equation}
p = p_\gamma + p_v \tag{5.9}
\end{equation}
and the time dependence is more complicated.
$\newline$

{ \bf Problem 5.1.}  Calculate the Hubble parameter $H$ as a
function of time for the two cases described by Eqs. (5.3) and (5.5)
where $k$
 is zero and the density is either only from radiation or only from
matter.

$\newline$

 { \bf Problem 5.2.}  Show that if $k$
 is 0 and the energy and pressure are as described in Problem 4.1 for a
constant $w$ , then
\begin{equation}
R \propto t^{\frac{2}{3+3w}}. \notag
\end{equation}
Note that Eqs. (5.3) and (5.5) are obtained as special cases when
$w$ is 1/3 and 0.

$\newline$

 {\bf Problem 5.3.}  Show that if $k$
 is 0 and the energy density is all from matter and vacuum, with
$\rho_m$ and $\rho_v$ both positive, $\text{then}^{\mathbf{8}}$
\begin{equation}
R^{\frac{3}{2}} \propto sinh \sqrt{6 \pi G \rho_v} t. \notag
\end{equation}

\section{Critical Density}
  When $k$ is zero, the Friedmann equation (2.5) says the density is
\begin{equation}
\rho = \frac{ 3 H^{2}}{ 8 \pi G} .  \tag{6.1}
\end{equation}
This is called the critical density. Let
\begin{equation}
\Omega = \frac{\rho}{ 3 H^{2}/ 8 \pi G}. \tag{6.2}
\end{equation}
Then $\Omega$ is 1 if $k$
 is zero.  From the Friedmann equation we see
$\Omega$
 is less than 1 if
$k$
 is -1 and
$\Omega$
 is larger than 1 if
$k$
 is 1.  Also let
\begin{equation}
\Omega_\gamma = \frac{\rho_\gamma}{\rho} \Omega \notag
\end{equation}
\begin{equation}
\Omega_m = \frac{\rho_m}{\rho} \Omega \notag
\end{equation}
\begin{equation}
\Omega_v = \frac{\rho_v}{\rho} \Omega. \tag{6.3}
\end{equation}

 We can write the Friedmann equation (2.5) as
\begin{equation}
\frac{ \rho - 3 H^{2}/ 8 \pi G}{\rho} = \frac{ 3kR(t_1)^{2}}{8 \pi G
\rho R^{2}}.  \tag{6.4}
\end{equation}
The left side is the fraction of the energy density that is
different from the critical density.  We know from observations that
right now in our universe this fraction is small. From the right
side of the equation we can see how this fraction changes as a
function of $R$
 as the universe expands.  In particular, using Eq. (4.6), we see that in
the early universe when the energy density was mostly from
radiation, the fraction (6.4) increased in proportion to $R^{2}$.
 It grew very much larger as the universe expanded. Since it is small now, it must have
been very small at early times.  That begs for explanation.  The
standard explanation is that in the very early universe there was a
period of inflationary expansion when the energy density was mostly
from vacuum. Then, since vacuum energy density stays constant, the
fraction (6.4) decreased in proportion to $1/R^{2}.$ It became very
very small as $R$ grew exponentially, as described by Eq. (5.7).
$\newline$

{ \bf Problem 6.1. } Show that if the value of
$\Omega$
 is 1 at a particular time, then it is constant in time.
$\newline$

{ \bf Problem 6.2. }  Can $\Omega_v$
 be constant in time?  What is required for that to happen?

\section{Acceleration}
  For the present and future expansion of the universe we
consider only matter and vacuum.  The energy density of radiation,
which was so important in the early universe, is negligible now.

  The acceleration of the expansion of the universe is
described by Eq. (3.4). With only $\rho_m$ and $\rho_v$ included, it
is
\begin{equation}
\frac{d^{2}R}{dt^{2}} = -\frac{4}{3} \pi G ( \rho_m - 2 \rho_v)R
\tag{7.1}
\end{equation}
because $p_m$ is zero and $p_v$ is $-\rho_v.$
  The acceleration is positive, zero, or negative depending on whether
$2 \rho_v$
 is greater than, equal to, or less than
$\rho_m$
 In the plane of
$\Omega_m$
 and
$\Omega_v$ , the acceleration is zero on the line where $2 \Omega_v
/ \Omega_m$
 is 1.  For larger
$2 \Omega_v / \Omega_m$ the acceleration is positive and for smaller
$2 \Omega_v / \Omega_m$
 the acceleration is negative. (See Fig.1)
$\newline$

{ \bf Problem 7.1.}  Suppose the acceleration Eq. (7.1)
is zero at a particular time.  Can it stay zero in time?  What is
required for that to happen?

\section{Expansion Forever?}

  Will the expansion of the universe ever stop?  That question
was easier to discuss a few years ago when only matter density was
considered.  If $\rho$ is just $\rho_m,$ we can use our initial
Newtonian equations.  Both the $M$ in Eqs. (2.2) and (2.3) and the
$E$ in Eqs. (2.2) and (2.4) are constant.  We can apply these
equations in the familiar Newtonian way.  Galaxies will stop moving
apart, so the expansion will stop, if $E$ is negative.  The
expansion will not stop if $E$ is zero or positive.

  Now we consider energy densities for both matter and vacuum.
We want to find the values of $\Omega_m$ and $\Omega_v,$ the area in
the plane of $\Omega_m$ and $\Omega_v$, for which the expansion will
stop.  We just found that when $\Omega_v$ is zero, the expansion
will stop if and only if $\Omega_m$ is larger than 1 ( because that
means $\Omega$ is larger than 1, $k$ is positive, and $E$ is
negative).
  The expansion will stop if $d^{2}R / dt ^{2}$ is negative and
remains negative until $dR/dt$ is zero.  We can see this happens if
$\Omega_v$ is negative, because then we see from Eq. (3.4) that

\begin{equation}
\frac{d^2R}{dt^2} \propto - (\Omega_m - 2 \Omega_v) \tag{8.1}
\end{equation}
is negative and will stay negative, and in Eq. (3.3)
\begin{equation}
\frac{8}{3} \pi G (\rho_m + \rho_v) R^2 \tag{8.2}
\end{equation}
will fall to $k R(t_1)^2$ even if $k$
 is -1, making
$dR/dt$
 zero, as
$\rho_m R^2$
 decreases in proportion to
$1/R$ and the magnitude of $\rho_v R^2$
 increases in proportion to
$R^2.$

  Suppose $\Omega_m \le 1.$ We have found that the expansion
stops if $\Omega_v$
 is negative but does not stop if
$\Omega_v$ is zero.  We can conclude that it does not stop if
$\Omega_v$
 is positive, because larger
$\Omega_v$ makes $d^{2}R / dt ^{2}$
 and
$dR / dt$
 both larger.

  For $\Omega_m > 1,$
 there is a similar limit on the
$\Omega_v$
 for which the expansion will stop.  The limit values of
$\Omega_v$
 as a function of
$\Omega_m$
 form a curve across the plane of
$\Omega_m$
 and
$\Omega_v$
 (see Fig. 1).  For smaller
$\Omega_v$
 the expansion will stop and for larger
$\Omega_v$
 it will not, again because larger
$\Omega_v$
 makes
$d^{2}R / dt ^{2}$
 and
$dR / dt$
 both larger.  We know that the expansion stops when
$\Omega_m$
 is larger than 1 and
$\Omega_v$
 is zero, so the limit values of
$\Omega_v$
 are at least zero.

 At the limit, $d^{2}R / dt ^{2}$
 is negative just long enough for
$dR / dt$
 to reach zero, so
$d^{2}R / dt ^{2}$
 and
$dR / dt$
 are zero at the same time.  We write
$\tilde{\rho}_m$
 and
$\tilde{R}$
 for the values at that time and
$\rho_m$
 and
$R$
 for the values now.  The value of
$\rho_v$
 is the same at all times.  From Eqs. (3.4), (3.3), and (4.4) we get

\begin{equation}
\tilde{\rho}_m - 2 \rho_v = 0 \tag{8.3}
\end{equation}
\begin{equation}
\frac{8}{3} \pi G (\tilde{\rho}_m + \rho_v) \tilde{R}^2 = k R(t_1)^2
= R(t_1)^2 \tag{8.4}
\end{equation}
\begin{equation}
\rho_m R^3 = \tilde{\rho}_m \tilde{R}^3 \tag{8.5}
\end{equation}
which imply
\begin{equation}
8 \pi G \rho_v (\frac{\rho_m}{2 \rho_v})^{\frac{2}{3}} R^2 =
R(t_1)^2. \tag{8.6}
\end{equation}
Using this to substitute for $R_0 ^2 / R^2$
 in the Friedmann equation (2.5), where
$k$
 must be 1, dividing by
$H^2$ and using the definitions (6.2) and (6.3) for $\Omega_m$
 and
$\Omega_v,$
 yields
\begin{equation}
1 - \Omega_m - \Omega_v = -\frac{3}{2^{\frac{2}{3}}} \Omega_v^{
\frac{1}{3} } \Omega_m^{\frac{2}{3}} \tag{8.7}
\end{equation}
which we can write as
\begin{equation}
4y^3 - 3y + x = 0 \tag{8.8}
\end{equation}
with
\begin{equation}
x = 1 - \frac{1}{\Omega_m}, \qquad  \qquad y = (\frac{\Omega_v}{4
\Omega_m})^{\frac{1}{3}}. \tag{8.9}
\end{equation}
Let $ y = cos \beta;$ that is allowed because $x$ and $y$ are
positive and Eq. (8.8) implies $y$ is less than 1. Using
\begin{equation}
cos 3 \beta = 4 cos^3 \beta - 3 cos \beta, \tag{8.10}
\end{equation}
and taking $y$ to be zero when $x$ is zero, so $\Omega_v$ is zero
when $\Omega_m$ is 1, gives
\begin{equation}
-x = cos 3 \beta \notag
\end{equation}
\begin{equation}
x = cos(3 \beta - \pi ) , \qquad  y = cos( \frac{1}{3} cos^{-1} x +
\frac{\pi}{3}) \tag{8.11}
\end{equation}
\begin{equation}
\Omega_v = 4 \Omega_m cos^3 (  \frac{1}{3} cos^{-1}(1 -
\frac{1}{\Omega_m}) + \frac{\pi}{3}). \tag{8.12}
\end{equation}

  On the curve in Fig.1, the values of $\Omega_v$
 are exaggerated by a factor of  5
to show that they are positive when $\Omega_m$
 is larger that 1. Actually they are quite small.
$\newline$

{\bf Problem 8.1.} For example, show that $\Omega_v /
\Omega_m$ is 1/54 when $\Omega_m$
 is $54/28$ on the curve described by Eq. (8.12), because if
$y$
 is 1/6 then
$x$ is 26/54 in Eq. (8.8).
$\newline$

{\bf Problem 8.2.} Show that a
good approximation to Eq. (8.12), obtained by dropping the first
term in Eq. (8.8), is
\begin{equation}
\Omega_v = \frac{4 \Omega_m}{27} (1 - \frac{1}{\Omega_m})^{3}.
\notag
\end{equation}

\section{Travel Time}

  We can calculate the relation between the redshift of light
from a distant galaxy and the time the light takes to reach us.  It
depends on the density of matter and vacuum energy.

 For light that was emitted from a distant galaxy at
wavelength $\lambda$
 and is observed here now at wavelength
$\lambda_0$ the redshift parameter $z$ is defined by
\begin{equation}
1 + z = \frac{\lambda_0}{\lambda}. \tag{9.1}
\end{equation}
As the universe expands, the wavelengths of light expand in
proportion to distances.  Let
\begin{equation}
r = \frac{R}{R_0} \tag{9.2}
\end{equation}
with $R$
 and
$r$ functions of time and $R_0$
 the value of
$R$ now.  Then at the time the light was emitted
\begin{equation}
r = \frac{1}{1+z}. \tag{9.3}
\end{equation}

 It is handy to work with $r$
 instead of
$R$
 and measure time in units of
$1 / H_0$ where $H_0$
 is the value of
$H$
 now.  Let
\begin{equation}
\tau = H_0 t, \tag{9.4}
\end{equation}
so $\tau$
 is 1 when
$t$
 is
$1/H_0.$
 Using Eq. (4.4), we can write
\begin{equation}
\rho_m = \rho_{m0} (\frac{R_0}{R})^3 = \frac{\rho_{m0}}{r^3}
\tag{9.5}
\end{equation}
where $\rho_{m0}$ is the value of $\rho_m$
 now.  Then using this, Eqs. (4.14), (6.2) and (6.3), we can write Eq. (7.1) as
\begin{equation}
\frac{d^2 r}{d \tau ^2} = -\frac{1}{2} \frac{\Omega_{m0}}{r^2} +
\Omega_{v0} r \tag{9.6}
\end{equation}
where $\Omega_{m0}$
 and
$\Omega_{v0}$
 are the values of
$\Omega_m$
 and
$\Omega_v$
 now. From the Hubble law now
\begin{equation}
\frac{dR}{dt} = H_0 R \tag{9.7}
\end{equation}
we get
\begin{equation}
\frac{dr}{d \tau} = r = 1 \tag{9.8}
\end{equation}
 at
$r = 1.$
 Remembering that Eq. (7.1) is a version of Eq. (3.4) which was obtained by
taking the time derivative of Eq. (3.3), we can deduce, and/or check
by differentiating again, that
\begin{equation}
(\frac{dr}{d \tau })^2 = \frac{\Omega_{m0}}{r} + \Omega_{v0} r^2 +
constant. \tag{9.9}
\end{equation}
Assuming that $dr/d \tau$
 is positive, and choosing the constant to get the value (9.8) when
$r$
 is 1, we obtain
\begin{equation}
\frac{d \tau }{dr} = (\frac{\Omega_{m0}}{r} + \Omega_{v0} r^2 + 1 -
\Omega_{m0} - \Omega_{v0} ) ^{- \frac{1}{2}}. \tag{9.10}
\end{equation}
We have not needed to know or assume anything about the constant
term in the Friedmann equation(3.3).

\qquad Integrating gives time, in units of $1/H_0,$
 as a function of the distance fraction
$r$
 or redshift
$z.$
 For each
$\Omega_{m0} , \Omega_{v0},$ the time the light takes to reach us is
\begin{equation}
(\Delta \tau)(r) = \int_r^{1} \frac{d \tau}{dr} (r') dr' \tag{9.11}
\end{equation}
with $d \tau / dr$
 given by Eq. (9.10). We can see qualitatively how this time depends on
$\Omega_{m0}$
 and
$\Omega_{v0}.$
 Since
$r$
 in the integral is less than 1, the
$\Omega_{m0} / r$ in Eq. (9.10) is larger than the $\Omega_{m0}$
 and
$\Omega_{v0} r^2$ is smaller than $\Omega_{v0},$
 so larger
$\Omega_{m0}$
 gives smaller time and larger
$\Omega_{v0}$
 gives larger
$\text{time.}^{\mathbf{9}}$

  The integrand (9.10) for $(\Delta \tau)(r)$ is valid as long
as matter and vacuum provide the dominant forms of energy density.
Radiation becomes important for $z$
 larger than about 10,000.  If there are no other forms of energy
density, extending the integral (9.11) to $(\Delta \tau) (0)$
 gives a very good approximation to the age of the universe for each
$\Omega_{m0}$ and $\Omega_{v0}.$

  Einstein's cosmological constant, which we now interpret as
vacuum energy, was introduced originally to get equations that can
describe a static universe.  That can not be described by the
equations we are using here.  In particular, Eq. (9.8) is clearly
inconsistent with a static universe;  the assumption we need to make
to use it is that the universe is expanding.  The way we are doing
things here leaves a piece of history behind.
$\newline$

{\bf Problem 9.1.}  Show that if there is only vacuum energy density
and $\Omega_{v0}$ is positive and less than 1, then
\begin{equation}
r = (\frac{1 - \Omega_{v0}}{\Omega_{v0}})^{\frac{1}{2}} sinh
\Omega_{v0}^{\frac{1}{2}} \tau. \notag
\end{equation}
Notice that $r \rightarrow \tau$ as $\Omega_{v0}$ approaches zero,
and the age of the universe becomes infinite as $\Omega_{v0}$
approaches 1.
$\newline$

{\bf Problem 9.2.}  Find how $r$ varies as a function of $\tau$ when
there is only vacuum energy density and $\Omega_{v0}$ is 1. Show
that then the age of the universe is infinite.
$\newline$

{\bf Problem 9.3.}  Show that if there is only vacuum energy density
and $\Omega_{v0}$ is larger than 1, then there is no big bang
because $dr/ d \tau$ does not stay positive for $r$ between zero and
1.
$\newline$

{\bf Problem 9.4.}  Suppose there is only matter and vacuum energy
density.  By looking at the minimum of $(dr / d \tau)^2$ as a
function of r, show that there is no big bang, because $dr/ d \tau$
does not stay positive for r between zero and 1, if
\begin{equation}
(2^{\frac{1}{3}} + 2^{- \frac{2}{3}}) \Omega_{m0}^{\frac{2}{3}}
\Omega_{v0} ^{\frac{1}{3}} + 1 - \Omega_{m0} - \Omega_{v0} \notag
\end{equation}
is not positive when $\Omega_{m0}$ is less  than $2 \Omega_{v0}.$
Sketch the curve that bounds the area in the $\Omega_{m0},
\Omega_{v0}$ plane where this happens.  Show that the age of the
universe grows infinitely  large as this area is $\text{approched.}
^{\mathbf{9}}$
$\newline$

{\bf Problem 9.5.}  Suppose there is only matter density and
$\Omega_{m0}$ is 1. Calculate $(\Delta \tau) (r)$ and the age of the
universe $(\Delta \tau) (0).$ Expand $(\Delta \tau) ((1+z)^{-1})$ as
a series in powers of $z$ to second order, which can be used for
small $z$ for $r$ near 1.
$\newline$

{\bf Problem 9.6.}  By changing the variable from $r \text{ to } z$
in the integral (9.11) and expanding in powers of $z,$ show that to
second order in $z, \text{ for small } z \text{ for } r$ near 1,
\begin{equation}
(\Delta \tau)( (1 + z)^{-1}) = z - z^2 - \frac{1}{4} ( \Omega_{m0} -
2  \notag \Omega_{v0}) z^2 .
\end{equation}
$\newline$

{\bf Problem 9.7.} Sketch the age of the universe $(\Delta \tau)(0)
\text{ as a function of } \Omega_{m0}$ $\text{for } \Omega_{m0}$
between 0 and 1 for the cases where: \newline (a) there is only
matter energy density; and \newline (b) there is both matter and
vacuum, $\Omega_{m0} \text{ and } \Omega_{v0} $ are both positive,
and their sum is 1. \newline The relative positions of the curves
and the directions of their slopes can be deduced from the
observations made following equation Eq. (9.11). The values when
$\Omega_{m0}$ is 1 are found in Problem 9.5.  For (b), the behavior
as $\Omega_{m0}$ approaces 0 is found in Problems 9.1 and 9.2.  For
(a), the value when $\Omega_{m0}$ is 0 can be deduced by seeing that
$d \tau / dr$ is 1.
$\newline$

{\bf Problem 9.8.}  Suppose there is only radiation energy density
and $\Omega_{\gamma 0 },$ the value of $\Omega_\gamma$ now, is 1.
>From Eq. (5.3) and the requirement that $d \tau / dr$ is 1 when $r$
is 1 deduce that
\begin{equation}
(\Delta \tau)(r) = \frac{1}{2} ( 1 - r^2). \notag
\end{equation}
In particular, the age of the universe in units of $1/H_0$ is
\begin{equation}
(\Delta \tau) (0) = \frac{1}{2}. \notag
\end{equation}
$\newline$

{\bf Problem 9.9.}  Show that if there were only radiation energy
density, Eq. (9.10) would be replaced by
\begin{equation}
\frac{ d \tau}{dr} = ( \frac{\Omega_{\gamma 0 } }{r^2} + 1 -
\Omega_{\gamma 0} ) ^{- \frac{1}{2}} \notag
\end{equation}
with $\Omega_{\gamma 0} > 0 $ the value of $\Omega_{\gamma}$ now.
Show that this gives times $(\Delta \tau)(r)$ that are smaller for
larger $\Omega_{\gamma 0 } \text{ and for each } \Omega_{\gamma 0 }$
are smaller than they would be if there were only matter energy
density with $\Omega_{m0} \text{ the same as } \Omega_{\gamma 0}.$
Sketch the age of the universe $(\Delta \tau)(0)$ as a function of
$\Omega_{\gamma 0 } \text{ for } \Omega_{\gamma 0 }$ between 0 and
1. (Find the value when $\Omega_{\gamma 0 }$
 is 0 by seeing that
$d \tau / dr$ is 1 and use the result of Problem 9.8 for the value
when $\Omega_{\gamma 0}$ is 1.) Compare this sketch with the
sketches made in Problem 9.7. Show, by calculating the derivative
with respect to $r$ and checking the value when $r$ is 1, that
\begin{equation}
(\Delta \tau)(r) = \frac{1 - \sqrt{ \Omega_{\gamma 0} + (1 -
\Omega_{\gamma 0})r^2}}{1 - \Omega_{\gamma 0 }}. \notag
\end{equation}
Show that in particular
\begin{equation}
(\Delta \tau)(0) = \frac{1}{1 + \sqrt{\Omega_{\gamma 0 }}} \notag
\end{equation}
and check that this agrees with your sketch.
$\newline$

{\bf Problem 9.10.}  Suppose the energy and pressure are as
described in Problems 4.1 and 5.2 with constant $w,$
 but $w$ not -1, and
$\Omega_{w 0},$ the present ratio of this ``$w$" energy density to
the critical density, is 1.  From the result of Problem 5.2 and the
requirement that $d \tau / dr$ is 1 when $r$ is 1 deduce that
\begin{equation}
(\Delta \tau)(r) = \frac{2}{3 + 3 w} (1 - r^{\frac{3 + 3 w}{2}}).
\notag
\end{equation}
In particular, the age of the universe in units of $1/H_0$ is
\begin{equation}
(\Delta \tau)(0) = \frac{2}{3 + 3 w}. \notag
\end{equation}
Check that you get the result of Problem 9.5 when $w$ is 0 and of
Problem 9.8 when $w$ is 1/3.
$\newline$

{\bf Problem 9.11.} Show that if the vacuum energy and pressure are
replaced by the $w$ energy and pressure  described in Problems 4.1,
5.2 and 9.10, for a constant $w,$ then Eq. (9.10) is replaced by
\begin{equation}
\frac{d \tau}{dr} = ( \Omega_{m0} r^{-1} + \Omega_{w0} r^{-1 - 3w} +
1 - \Omega_{m0} - \Omega_{w0} ) ^{-\frac{1}{2}} \notag
\end{equation}
where $\Omega_{w0}$ is the present ratio of the $w$ energy density
to the critical density.  Note that when $w$ is -1/3 this and the
travel times $(\Delta \tau)(r)$ are the same as for matter alone.
Check that you get the expected results when $w$ is 0 or -1. Show
that if $w$ is less than -1/3, the times $(\Delta \tau)(r)$ are
larger for larger $\Omega_{w0};$ in particular, they are larger when
$\Omega_{w0}$ is positive than when $\Omega_{w0}$ is 0.  Suppose
$\Omega_{m0} \text{ and } \Omega_{w0}$ are both positive, their sum
is 1, and $w$ is -1/2.  Sketch the age of the universe $(\Delta
\tau)(0)$ as a function of $\Omega_{m0} \text{ for } \Omega_{m0}$
between 0 and 1, as in Problem 9.7, taking the value when
$\Omega_{m0}$ is 1 from Problem 9.5 and the value when $\Omega_{m0}$
is 0 from Problem 9.10.  By calculating the derivative with respect
to $r$ and checking the value when $r$ is 1, show that
\begin{equation}
(\Delta \tau) (r) = \frac{4}{3 \Omega_{w0}} ( 1 - \sqrt{\Omega_{m0}
+ \Omega_{w0} r^{\frac{3}{2}}}) . \notag
\end{equation}
Show that in particular
\begin{equation}
(\Delta \tau)(0) = \frac{4}{3(1 + \sqrt{\Omega_{m0}})}. \notag
\end{equation}
Check that this agrees with your sketch and compare it with the
result of Problem 9.9 for radiation.
$\newline$

{\bf Problem 9.12.}
By calculating the derivative with respect to $r$ and checking the
value when $r$ is 1,  show that if $\Omega_{m0} \text{ and }
\Omega_{v0}$ are both positive and their sum is 1,
$\text{then}^{\mathbf{10}}$
\begin{equation}
(\Delta \tau)(r) = \frac{2}{3} \frac{1}{\sqrt{\Omega_{v0}}} [
sinh^{-1} \sqrt{\frac{\Omega_{v0}}{\Omega_{m0}}} -  sinh^{-1}
\sqrt{\frac{\Omega_{v0}}{\Omega_{m0}}} r^{\frac{3}{2}}]. \notag
\end{equation}
In particular, the age of the universe in units of $1/H_0$ as a
function of $\Omega_{m0} \text{ for } \Omega_{m0}$ between 0 and 1
is
\begin{equation}
(\Delta \tau) (0) = \frac{2}{3} (1 - \Omega_{m0})^{-\frac{1}{2}}
sinh^{-1} (\frac{1}{\Omega_{m0}} - 1)^{\frac{1}{2}}. \notag
\end{equation}
Check that this agrees with the sketch you made for Problem 9.7(b).

$\newline$
{\bf Problem 9.13.}  Suppose there is only matter density
and $\Omega_{m0}$ is positive and less than 1. By calculating the
derivative with respect to $r$ and checking the value when $r$ is 1,
show $\text{that} ^{\mathbf{11}}$
\begin{equation}
(\Delta \tau) (r) = \frac{ 1 - r \sqrt{    \Omega_{m0} / r + 1 -
\Omega_{m0}    }    }       {1 - \Omega_{m0}} \qquad \qquad \qquad
\qquad \qquad \qquad \qquad \notag
\end{equation}
\begin{equation}
 + \frac{\Omega_{m0}}{2(1-\Omega_{m0})^{\frac{3}{2}}} [cosh^{-1} (1 + 2
\frac{1 - \Omega_{m0}}{\Omega_{m0}}r) - cosh^{-1} (1 + 2\frac{1 -
\Omega_{m0}}{\Omega_{m0}})]. \notag
\end{equation}
In particular, the age of the universe in units of $1/H_0$ as a
function of $\Omega_{m0} \text{ for } \Omega_{m0} $ between 0 and 1
is
\begin{equation}
(\Delta \tau) (0) = \frac{1}{1 - \Omega_{m0}} -
\frac{\Omega_{m0}}{2(1 - \Omega_{m0})^{\frac{3}{2}}} cosh^{-1} ( 1 +
2 \frac{1 - \Omega_{m0}}{\Omega_{m0}}). \notag
\end{equation}
Check that this agrees with the sketch you made for Problem 9.7(a).

\section{Hubble Plots}

  Hubble plots are providing a more detailed picture of the
universe as they extend to larger distances.  Earlier plots showed
galaxies receding from us with velocities proportional to their
distances, as expected in a uniformly expanding universe.  Now
Hubble plots are showing how the expansion has developed during the
time light from distant galaxies has been traveling to
$\text{us.}^{\mathbf{12}}$

Recession velocities are measured from red shifts of spectral lines.
Hubble plots now show distance as a function of red shift.  Velocity
is considered to be a secondary quantity that can be calculated from
the formula for the relativistic Doppler effect.

The distance to an object is found by comparing its apparent
luminosity, how bright it looks, with its absolute luminosity, how
bright it is thought to actually be.  Apparent luminosity decreases
with distance, but the expansion of the distance scale, the
curvature of space, and the redshift all combine to make the
distance used to describe luminosity different from the speed of
light times the travel time.

If the redshift is $z,$ the light from the object is spread over a
sphere of radius
\begin{equation}
\chi( r = (1 + z)^{-1}) = \int_r ^1 \frac{1}{r'} \frac{d \tau}{dr'}
(r') dr' \tag{10.1}
\end{equation}
in units of $1/H_0,$ because distances traveled at earlier times are
expanded by factors of $1/r'.$ If space is flat, has no curvature,
the light is spread over a sphere of area $4 \pi (  \chi / H_0)^2$.
General relativity says space may be curved, so the area of the
sphere of light is $4 \pi ( \sigma / H_0)^2$ where $\sigma \text{ is
} \chi$ if there is no curvature, but
\begin{equation}
\sigma = r_c sin(\frac{\chi}{r_c}) \tag{10.2}
\end{equation}
for positive curvature, and
\begin{equation}
\sigma = r_c sinh(\frac{\chi}{r_c}) \tag{10.3}
\end{equation}
for negative curvature, with
\begin{equation}
r_c = | 1 - \Omega_0 | ^{-\frac{1}{2}} \tag{10.4}
\end{equation}
where $\Omega_0 = \Omega_{m0} + \Omega_{v0}$ is the value of $\Omega
\text{ now.}^{\mathbf{13}}$ The two-dimensional analog that is easy
to picture is that on the two-dimensional surface of a sphere of
radius $r_c$ the circumference of a circle of radius $\chi \text{ is
} 2 \pi r_c sin( \chi / r_c).$

The redshift decreases the frequencies of the light and the energies
of the photons by a factor of $(1+z)^{-1},$ and the frequencies of
arrival of photons, the numbers of photons arriving per unit time,
are decreased by the same factor, so the redshift decreases the
intensity of the light by a factor of $(1+z)^{-2}.$ The ``luminosity
distance" $d_L$ that is used in Hubble plots is defined by
\begin{equation}
d_L = (1+z) \frac{\sigma}{H_0} \tag{10.5}
\end{equation}
so that
\begin{equation}
\frac{1}{4 \pi d_L ^2} = \frac{1}{ (1+z)^2 4 \pi ( \sigma / H_0
)^2}. \tag{10.6}
\end{equation}
It is what we would deduce the distance to the object to be if the
relation between its apparent luminosity and absolute luminosity
came just from the light spreading out over a sphere in flat space.
If there is only matter density, the integral (10.1) with Eq. (9.10)
for $d \tau / dr$ yields the simple result
$\text{that}^{\mathbf{14}}$
\begin{equation}
d_L= \frac{2}{H_0 \Omega_{m0} ^2} [ \Omega_{m0} z + (2 -
\Omega_{m0})( 1 - \sqrt{ \Omega_{m0} z + 1})] \tag{10.7}
\end{equation}
for all three cases of positive, zero, or negative curvature where
$\Omega_{m0}$ is greater than, equal to, or less than 1.
$\newline$

{\bf Problem 10.1.}  Show that Eq. (10.7) holds when there is only
matter density, no curvature, and $\Omega_{m0}$ is 1.
$\newline$

{\bf Problem 10.2.} Suppose there is only matter density and
$\Omega_{m0}$ is larger than 1.  By calculating the derivative with
respect to $r$ and checking the value when $r$ is 1, show that
\begin{equation}
\chi(r) = \frac{2}{\sqrt{\Omega_{m0} -1}} [ sin^{-1}
\sqrt{\frac{\Omega_{m0} -1}{\Omega_{m0}}} - sin^{-1}
\sqrt{\frac{\Omega_{m0} -1}{\Omega_{m0}}r}]. \notag
\end{equation}
Show that this and Eqs. (10.2), (10.4) and (10.5) give the result
(10.7) for the luminosity distance $d_L.$ (The case where
$\Omega_{m0}$ is less than 1 is covered in Problem 10.9.)
 $\newline$

{\bf Problem 10.3.}  By changing the variable from $r \text{ to } z$
in the integral (10.1), using Eq. (9.10) for $d \tau / dr,$ and
expanding  in powers of $z,$ show that to second order in $z, \text{
for small } z \text{ for } r$ near 1,
\begin{equation}
H_0 d_L = z + \frac{1}{2} z^2 - \frac{1}{4}(\Omega_{m0} - 2
\Omega_{v0}) z^2 \notag
\end{equation}
for all three cases of positive, zero, or negative curvature for any
$\Omega_{m0} \text{ and } \Omega_{v0}.$
 $\newline$

{\bf Problem 10.4.}  Suppose $\Omega_{m0} \text{ and } \Omega_{v0}$
are both positive and their sum is 1.  By making observations like
those made for travel times following Eq. (9.11), show that for each
$z$ the luminosity distance $d_L$ is smaller for larger $\Omega_{m0}
\text{ and smaller } \Omega_{v0}.$
$\newline$

{\bf Problem 10.5.}  Suppose there is only matter with $\Omega_{m0}$
positive and less than 1. Show, using Eqs. (9.10), (10.1) and
(10.3)-(10.5), that for each $z$ the luminosity distance $d_L$
 is smaller for larger
$\Omega_{m0}.$

$\newline$

{\bf Problem 10.6.}  Suppose vacuum is replaced by $w$ energy and
pressure as in Problem 9.11. Suppose $\Omega_{m0} \text{ and }
\Omega_{w0}$ are both positive and their sum is 1.  Show that for
each $z \text{ and fixed } \Omega_{m0}$ the luminosity distance
$d_L$ is smaller for larger $w,$ and for each $z$ and fixed negative
$w,$ the luminosity distance $d_L$ is smaller for larger
$\Omega_{m0}.$
$\newline$

{\bf Problem 10.7.}  Suppose there is only $w$ energy and pressure
as described in Problems 4.1, 5.2, 9.10 and 9.11, and suppose
$\Omega_{w0}$ is 1. Show that if $w$ is not -1/3 then
\begin{equation}
H_0 d_L = \frac{2}{3w +1} (1+z) [ 1 - (1+z)^{-\frac{3w + 1}{2}}].
\notag
\end{equation}
$\newline$

{\bf Problem 10.8.}  Suppose vacuum is replaced by $w$ energy and
pressure as in Problems 9.11 and 10.6.  Now let $w$ be -1/3. Suppose
$\Omega_{m0} \text{ and } \Omega_{w0}$ are both positive and their
sum is 1. Then the luminosity distance is
\begin{equation}
d_L = (1+z) \frac{\chi(r)}{H_0} \notag
\end{equation}
with $\chi(r)$ given by Eqs. (10.1) and (9.10). Show, by calculating
the derivative with respect to $r$ and checking the value when $r$
is 1, that
\begin{equation}
\chi(r) = \frac{2}{\sqrt{1 - \Omega_{m0} }} [ sinh^{-1}
\sqrt{\frac{1 - \Omega_{m0}}{\Omega_{m0}}} - sinh^{-1} \sqrt{\frac{1
- \Omega_{m0}}{\Omega_{m0}}r}]. \notag
\end{equation}
Observe that $\chi(r)$ would be the same if there were only matter,
with the same $\Omega_{m0},$ but show, using Eq. (10.3), that the
luminosity distance $d_L$ would be larger.
$\newline$

{\bf Problem 10.9.}  Suppose there is only matter, with
$\Omega_{m0}$ positive and less than 1.  Show that Eqs. (10.3) -
(10.5) and the result of Problem 10.8 for $\chi(r)$ give the formula
(10.7) for the luminosity distance $d_L.$

\pagebreak
\section*{APPENDIX A. RADIATION PRESSURE}

 Consider the pressure on a small plane area $A$ from photons of
momentum $\vec{p}$ reflecting off it.  Let $p_x$ be the component of
$\vec{p}$ perpendicular to $A$

for a photon approaching $A.$ Reflection changes $p_x \text{ to }
-p_x$ and leaves the other components of $\vec{p}$ unchanged.  Force
is the rate of change of momentum, so the force on $A \text{ is } 2
p_x$ times the number of photons that hit $A$ per unit time.
Altogether, including photons with different momenta $\vec{p},$ the
force on $A$ is
\begin{equation}
\frac{1}{2} 2 \overline{ p_x v_x } A n_\gamma \tag{A.1}
\end{equation}
where $v_x$ is the $x$ component of the velocity of a photon with
momentum $\vec{p} \text{ and } n_\gamma $ is the number of photons
per unit volume.  The average is over all $\vec{p},$ but only
photons with positive $p_x$ exert force on $A.$ They are just half
of the photons present.  That explains the factor 1/2.  The velocity
of a photon is $c$ in the direction of its momentum.  We assume
there is a uniform distribution of momentum directions.  Then the
pressure is
\begin{equation}
p_\gamma = \overline{p_x v_x} n_\gamma = \overline{ p_x ( p_x / |
\vec{p} | ) } c n_\gamma = c \frac{1}{3} \frac{ \overline{ p_x ^2 +
p_y ^2 + p_z ^2}}{| \vec{p}|} n_\gamma \notag
\end{equation}
\begin{equation}
= \frac{1}{3} c \overline{| \vec{p} |} n_\gamma = \frac{1}{3}
\rho_\gamma \tag{A.2}
\end{equation}
because the average does not depend on the direction of the momentum
and the energy of a photon is
\begin{equation}
h \nu = c \frac{h}{\lambda} = c | \vec{p} | .\tag{A.3}
\end{equation}

\section*{APPENDIX B. VACUUM PRESSURE}

 The energy density is the time-time component $T_{00}$ of a relativistic
tensor $T_{\mu \nu}.$ That tells us how it is changed by Lorentz
transformations.  It is more complicated than the density of
something that is Lorentz invariant, for example electric charge. If
an observer sees electric charges at rest occupying a volume $V$
with charge density $\rho_e,$ an observer who sees the charges
moving with velocity $\vec{v}$ will see the same total charge
$\rho_e V$ occupying volume $V \sqrt{1 - \vec{v} ^2}$ with density
$\rho_e / \sqrt{1 - \vec{v}^2}.$ Charge density is the time
component $j_0$ of a relativistic four-vector
\begin{equation}
j_\mu = ( \frac{\rho_e}{\sqrt{1 - \vec{v}^2}} , \frac{\rho_e
\vec{v}}{\sqrt{1 - \vec{v}^2}}). \tag{B.1}
\end{equation}
The space part $\vec{j}$
 is charge per unit volume times velocity.  It is the current density.
It gives the charge crossing unit area per unit time.

The energy of an object is not Lorentz invariant.  If an observer
sees the object at rest with energy $m,$ an observer who sees it
moving with velocity $\vec{v}$ will say it has energy $m / \sqrt{ 1
- \vec{v}^2}.$ Its energy is the time component $p_0$ of a
relativistic four-vector
\begin{equation}
p_\mu = ( \frac{m}{\sqrt{1 - \vec{v}^2}} , \frac{m \vec{v}}{\sqrt{1
- \vec{v}^2}}). \tag{B.2}
\end{equation}
The space part $\vec{p}$ is the object's momentum.  Energy density
is the time-time component $T_{00}$ of a relativistic tensor $T_{\mu
\nu} \text{ like } p_\mu p_\nu .$ The other $T_{0 \nu}$ components
are momentum density or energy current.  The space-space components
$T_{jk},$ where $j \text{ and } k$  are 1,2, or 3, are momentum
current.  They give momentum per unit time crossing unit area.  That
is pressure.

In a homogeneous uniformly expanding universe, for observers who see
the nearby universe at rest, $T_{\mu \nu}$ must be zero for $\mu
\neq \nu.$ There is no momentum or flow of energy, and no shear
pressure (for example, no flow in the x direction of momentum in the
y direction).  Also
\begin{equation}
T_{11} = T_{22} = T_{33} ; \tag{B.3}
\end{equation}
if not, a coordinate rotation would make $T_{jk}$
 nonzero for some
$j$ different from $k.$ So

\begin{equation}
T_{\mu \nu} =   \left( \begin {array}{cccc} \rho&0&0&0 \\
\noalign{\medskip} 0&p&0&0  \\ \noalign{\medskip} 0&0&p&0 \\
\noalign{\medskip} 0&0&0&p \end {array} \right)  .  \tag{B.4}
\end{equation}
In a homogeneous universe, $\rho \text{ and } p$ must be the same
throughout space.  They can be functions only of time.

A Lorentz transformation changes $T_{\mu \nu}$ to
\begin{equation}
T_{\mu \nu} ' = \sum_{\alpha = 0} ^3 \sum_{\beta = 0} ^3
\Lambda_{\mu \alpha} \Lambda_{\nu \beta} T_{\alpha \beta} \tag{B.5}
\end{equation}
where $\Lambda$ is the 4$\times$4 matrix for the Lorentz
transformation of space-time coordinates.  For an observer moving
with velocity $-v$ in the $z$ direction, for example,
\begin{equation}
\Lambda =   \left( \begin {array}{cccc} \frac{1}{\sqrt{1 -
v^2}}&0&0&\frac{v}{\sqrt{1 - v^2}} \\ \noalign{\medskip} 0&1&0&0  \\
\noalign{\medskip} 0&0&1&0 \\ \noalign{\medskip} \frac{v}{\sqrt{1 -
v^2}}&0&0&\frac{1}{\sqrt{1 - v^2}} \end {array} \right)    \tag{B.6}
\end{equation}
gives

\begin{equation}
T_{00}' = \frac{\rho + v^2p}{1 - v^2} \notag
\end{equation}
\begin{equation}
T_{33}' = \frac{p + v^2\rho}{1 - v^2}. \tag{B.7}
\end{equation}
The energy-momentum tensor is Lorentz invariant, $T_{\mu \nu} '$ is
the same as $T_{\mu \nu},$ if and only if
\begin{equation}
p = - \rho.  \tag{B.8}
\end{equation}
If the vacuum is Lorentz invariant, then $p_v \text{ is } - \rho_v.$

\begin{figure}
\begin{center}
\epsfig{file=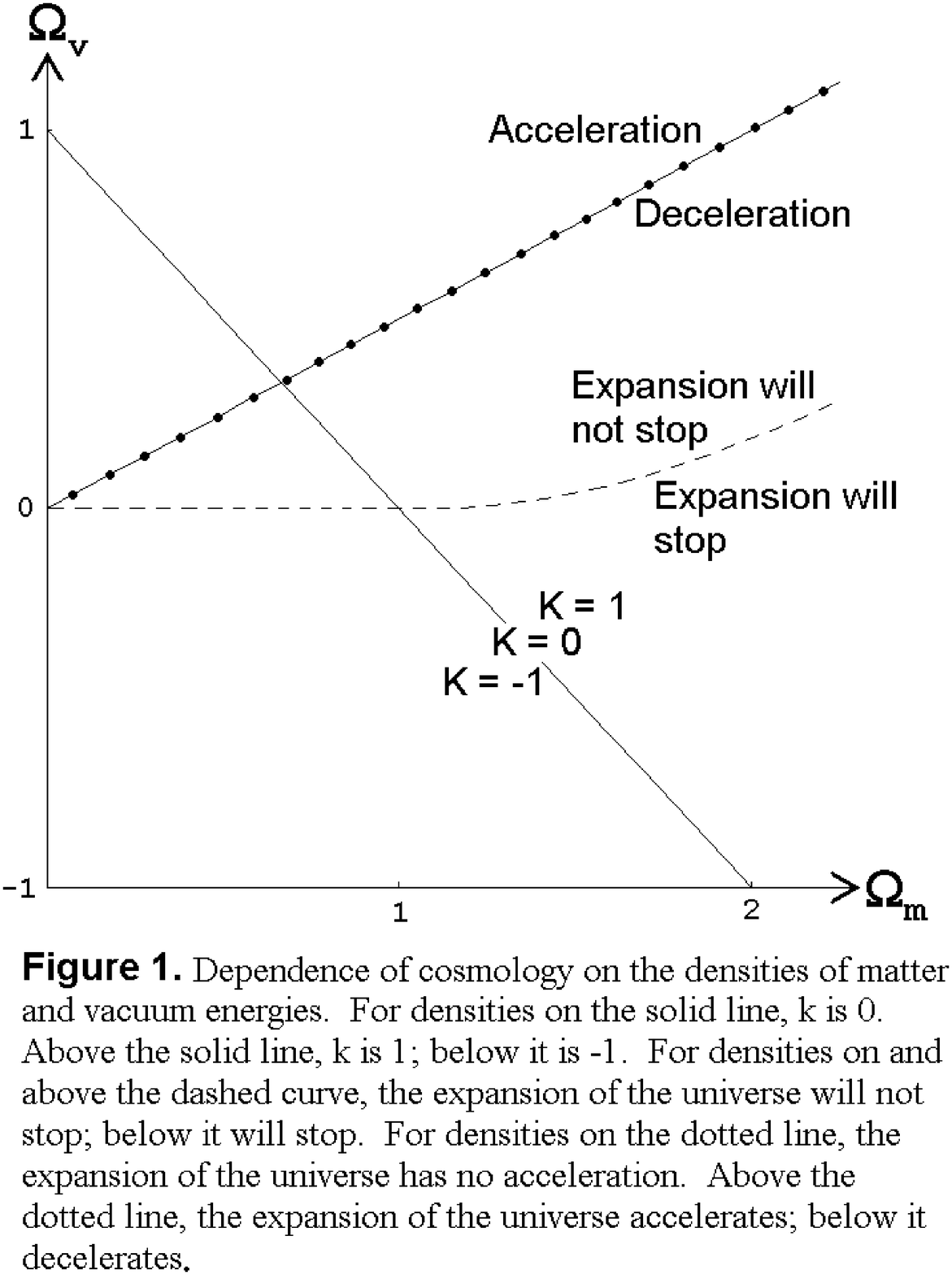, width=12cm}
\end{center}
\end{figure}

\pagebreak

\end{document}